\documentclass[preprint,showpacs,preprintnumbers,amsmath,amssymb]{revtex4}

\usepackage{graphicx}
\usepackage{dcolumn}
\usepackage{bm}

\begin{document}

\preprint{APS/123-QED}

\title{Large-scale intermittency of liquid-metal channel flow in a magnetic field}

\author{Thomas Boeck}
\author{Dmitry Krasnov}
\author{Andr\'e Thess}
\affiliation{Fakult\"at f\"ur Maschinenbau, Technische Universit\"at Ilmenau,\\
Postfach 100565, 98684 Ilmenau, Germany}
\author{Oleg Zikanov}
\affiliation{Dept. of Mechanical Engineering, University of Michigan-Dearborn, 48128-1491 MI, USA }

\date{\today}

\begin{abstract}
We predict a novel flow regime in liquid metals under the
influence of a magnetic field.  It is characterised
by long periods of nearly steady, two-dimensional flow interrupted
by violent three-dimensional bursts. Our prediction has been
obtained from direct numerical simulations in a channel geometry
at low magnetic
Reynolds number and translates into physical parameters which are
amenable to experimental verification under laboratory conditions.
The new regime occurs in a wide range of parameters and may have
implications for metallurgical applications.
\end{abstract}


\maketitle Liquid metals interact with magnetic fields under various
 circumstances ranging from electromagnetic flow
control \cite{Davidson:1999} and electromagnetic flow
measurement \cite{Shercliff:1962,Thess:2006} to the
generation of Earth's magnetic field \cite{Moffatt:1978} and
the laboratory studies of the magnetorotational instability \cite{Stefani:2006}. It is
widely believed, especially in the community dealing with
metallurgical applications, that a magnetic
field always damps turbulence and helps to reduce undesired
velocity fluctuations. In the present Letter we show that this
view is an oversimplification not always agreeing with reality.
We predict a novel flow regime, referred
to as large-scale intermittency (LSI), where the application of a
magnetic field to the flow of a liquid metal in a channel leads 
to repeating violent transitions between two-dimensional (2D) states, 
in which turbulence is fully suppressed, 
and fully turbulent three-dimensional (3D) states.
Similar intermittent dynamics was detected in two earlier studies 
of highly idealized flows: forced turbulence in a periodic box \cite{Zikanov:1998}
and inviscid flow in a tri-axial ellipsoid \cite{Thess:2007}.  
The channel configuration 
considered in the present Letter is the first, 
in which realistic flow conditions are approached 
by taking into account the effects of solid walls, 
viscosity, and mean shear.

In the following, we assume that the magnetic Reynolds number $Re_m$ 
is small, which applies to practically all industrial and laboratory 
flows of liquid metals. This allows us to employ the quasi-static 
approximation, whereby the induced magnetic field is negligibly small 
in comparison with the imposed field and adjusts instantaneously to 
the velocity fluctuations.  

An obvious effect of a static magnetic field on the flow of a liquid metal
is Joule dissipation of the induced currents, which provides
an additional mechanism of flow
suppression by  conversion of its kinetic energy  into heat.
Moreover, the flow can become
anisotropic or even 2D. This can be seen from 
the rate of Joule dissipation of a Fourier velocity mode
$\hat{\bm{u}}(\bm{k},t)$, which is $\mu(\bm{k})=\sigma
B^2\rho^{-1}|\hat{\bm{u}}|^2\cos^2\alpha$, where $\alpha$ is the angle
between the imposed
magnetic field $\bm{B}$ and the wavenumber vector $\bm{k}$, $\sigma$
is the conductivity and $\rho$ is the density of the liquid.
Proportionality to $\cos^2\alpha$ means that $\mu$ increases from
zero at $\bm{k}\bot \bm{B}$ to the maximum for modes with
$\bm{k}\|\bm{B}$. The magnetic field tends to eliminate velocity
gradients and elongate the flow structures in the direction of the
magnetic field lines.  The flow becomes axisymmetrically anisotropic
or, if the magnetic field is sufficiently strong, 2D
with all variables uniform in the direction of the magnetic
field \cite{Moffatt:1967}. 
 Similar anisotropic behavior has also been noted for 
magnetohydrodynamic 
turbulence with higher $Re_m$, in particular with magnetic Prandtl numbers 
$P_m=Re_m/Re\sim 1$
\cite{Montgomery:1981,Bigot:2008}.

Without reference to a specific flow geometry, the
LSI may evolve according to the following
scenario.  Under the action of the magnetic field, an initially
3D flow evolves into a pattern of nearly 
2D structures. The flow gradients along the magnetic
field are very weak in this state, so the Joule dissipation
decreases to nearly zero.  If, however, the 2D state is
not a stable attractor of the Navier-Stokes equations and the
magnetic field is not strong enough to completely suppress 3D
instabilities, perturbations grow and destroy the 2D structures.
The flow enters a 3D turbulent state and the process repeats itself.
This scenario has far-reaching implications for specific flows and
for low-$Re_m$ magnetohydrodynamics (MHD) in general.
The flows acquire
properties unforeseeable under statistical equilibrium
assumptions for MHD turbulence, whereby
the flow is either
nearly isotropic, statistically steady anisotropic, or 2D
depending on the strength of the magnetic field \cite{Alemany:1979}.
The existence of intermittent regimes relates
to the fundamental question of realizability of purely
2D states under the action of a magnetic field \cite{Thess:2007}.
The phenomenon is also of interest for general theory of hydrodynamic
instability, bifurcations, and transition to turbulence in parallel
shear flows. 
The effect of magnetic
field leads to new, unexpected roles of spanwise Tollmien-Schlichting (TS)
modes and streamwise streaks, the main agents of transition in 
ordinary hydrodynamics \cite{Schmid:Henningson:2001}. 

In the present Letter  we consider pressure-driven flow in a plane channel.
The imposed magnetic field is uniform and oriented in the  spanwise
direction, i.e. parallel to the walls and orthogonal to the flow.
Non-dimensional governing equations and boundary conditions for the
velocity  $\bm{u}$ and the electric potential $\phi$ are
\begin{eqnarray}
\label{MHDeqn}\nonumber && \frac{\partial \bm{u}}{\partial t} +
(\bm{u}\cdot \nabla)\bm{u} = -\nabla p+\frac{1}{Re}\nabla^2 \bm{u} +
\frac{Ha^2}{Re}\left(\bm{j}\times \bm{e}_y\right)\\
\nonumber && \bm{j}= -\nabla \phi   + \bm{u}\times \bm{e}_y,
\quad \nabla^2 \phi =\nabla \cdot\left(\bm{u} \times \bm{e}_y\right),\\
\nonumber & & \nabla \cdot\bm{u}  =  0, \quad u = v = w =
\frac{\partial \phi}{\partial z}=0 \hskip3mm\mbox{at } z=\pm 1,
\end{eqnarray}
where $x$, $y$, and $z$ coordinates are in the streamwise, spanwise,
and wall-normal directions, respectively.  The 
parameters are the hydrodynamic Reynolds number $Re\equiv
UL\nu^{-1}$ and the Hartmann number $Ha\equiv B L (\sigma/\rho
\nu)^{1/2}$, with $L$ and $U$ being the half-width of the channel
and the centerline velocity of the Poiseuille parabolic velocity
profile. 

The problem is solved  for two computational domains
of dimensions $2\pi \times 4\pi \times 2$ and $8\pi \times 4\pi \times 2$ 
in the $x,y,z$-directions using direct numerical simulation (DNS)
with  a  Fourier-Chebyshev method  and periodic
boundary conditions for $x$ and $y$ \cite{Krasnov:2007}.
The numerical resolution is $64^3$ and $256\times 64^2$ 
collocation points for the small and large domains, respectively.
The 
volume flux  per span width remains constant during the computations,
which are conducted at 
$Re=8000$, i.e. above the threshold $Re_c\approx 5772$ of the linear
instability. The velocity perturbation is defined as
$\bm{u}'=\bm{u}-\langle\bm{u}\rangle$, where $\langle\bm{u}\rangle$
is the mean velocity obtained by horizontal  averaging over the 
computational domain. 
In the LSI cycle discussed below,
the amplitude of $\bm{u}'$ decreases to the level of machine
round-off. 
To remove the resulting ambiguity and to mimic the noise
in actual flows, white noise with the amplitude
$10^{-6}$ relative to the mean flow is added  at
every time step.  The initial conditions correspond to
purely 2D flow \cite{Jimenez:1990}. Other initial conditions 
produced identical behavior after transients. 

The spanwise magnetic field does not interact with the base flow
or with any other 2D flow uniform in the spanwise
direction. This, in particular, includes the spanwise-independent
TS modes of linear instability, which implies that the primary
linear instability (but not the secondary 3D breakdown
of TS modes) is insensitive to the presence of the magnetic field.
Above $Re_c$, the phase space for the nonmagnetic channel flow
contains an attractor of 3D turbulence and two unstable
equilibria: the Poiseuille solution and a 2D channel
flow solution, which  takes the form of a steady traveling
wave in the short domain and of a chaotic wavetrain in the 
longer domain\cite{Jimenez:1990}.
In the presence of magnetic field, the same states exist but,
as found in our computations, their stability and attraction
basins  change. When  the magnetic field is weak
($Ha\lesssim 40$ at $Re=8000$), a solution with arbitrary initial
conditions converges to a 3D turbulent state, which has pronounced
anisotropic properties considered elsewhere\cite{Lee:Choi:2001,Krasnov:2008}.
By contrast, at sufficiently strong magnetic fields
($Ha \gtrsim 160$ at $Re=8000$), the 2D channel flow
solution\cite{Jimenez:1990} becomes the only stable attractor.

\begin{figure}[ht]
\centerline{
\includegraphics[width=0.45\textwidth]{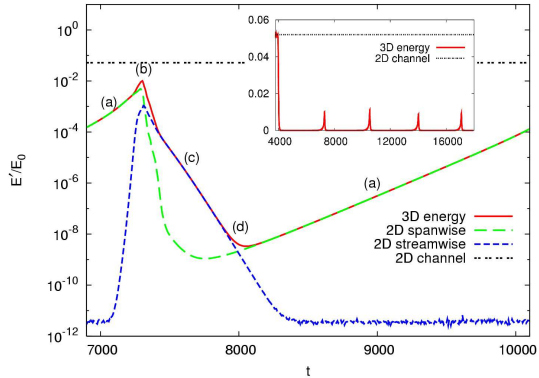}
} \caption{\label{fig1:re08_080_en}{Evolution of the perturbation energy
$E^{\prime}$
during one intermittency cycle. The total (3D) energy and the 
energy of 2D spanwise-independent and streamwise-independent perturbations are
depicted. For comparison, the perturbation energy of 2D channel
flow \cite{Jimenez:1990} is shown.  Letters \emph{a} to \emph{d}
indicate the flow stages illustrated in Fig. \ref{fig1:re08_080_zones}.
Time is nondimensional in convective units $L/U$.
Inset - evolution of perturbation energy during the entire run.}}
\end{figure}

The focus of this Letter is on  intermediate values of  $Ha$, for
which intermittency appears as 
a phase trajectory looping between base flow and turbulent state.
Results for $Ha=80$ and the short computational domain are presented, 
although intermittency with qualitatively similar basic
characteristics was observed at other intermediate values of $Ha$ and
in the longer domain.
The energy of velocity perturbations normalized by the
energy $E_0$ of the basic flow 
is shown in Fig. \ref{fig1:re08_080_en} as a function of
time.  As can be seen in the inset, soon after the start, 
the flow settles into an intermittent behavior. Long periods, 
during which the perturbations
are negligibly weak, are interrupted by short periods of strong
perturbations. The 2D channel flow solution \cite{Jimenez:1990}
is never approached.
 The intermittency events form a regular pattern with
approximately constant periods between the bursts and without
noticeable tendency for decay or growth of burst intensity.

\begin{figure}[ht]
\scriptsize{
\parbox{0.22\linewidth}{(a)}\parbox{0.76\linewidth}{(b)}
\centerline{
\includegraphics[width=0.23\textwidth]{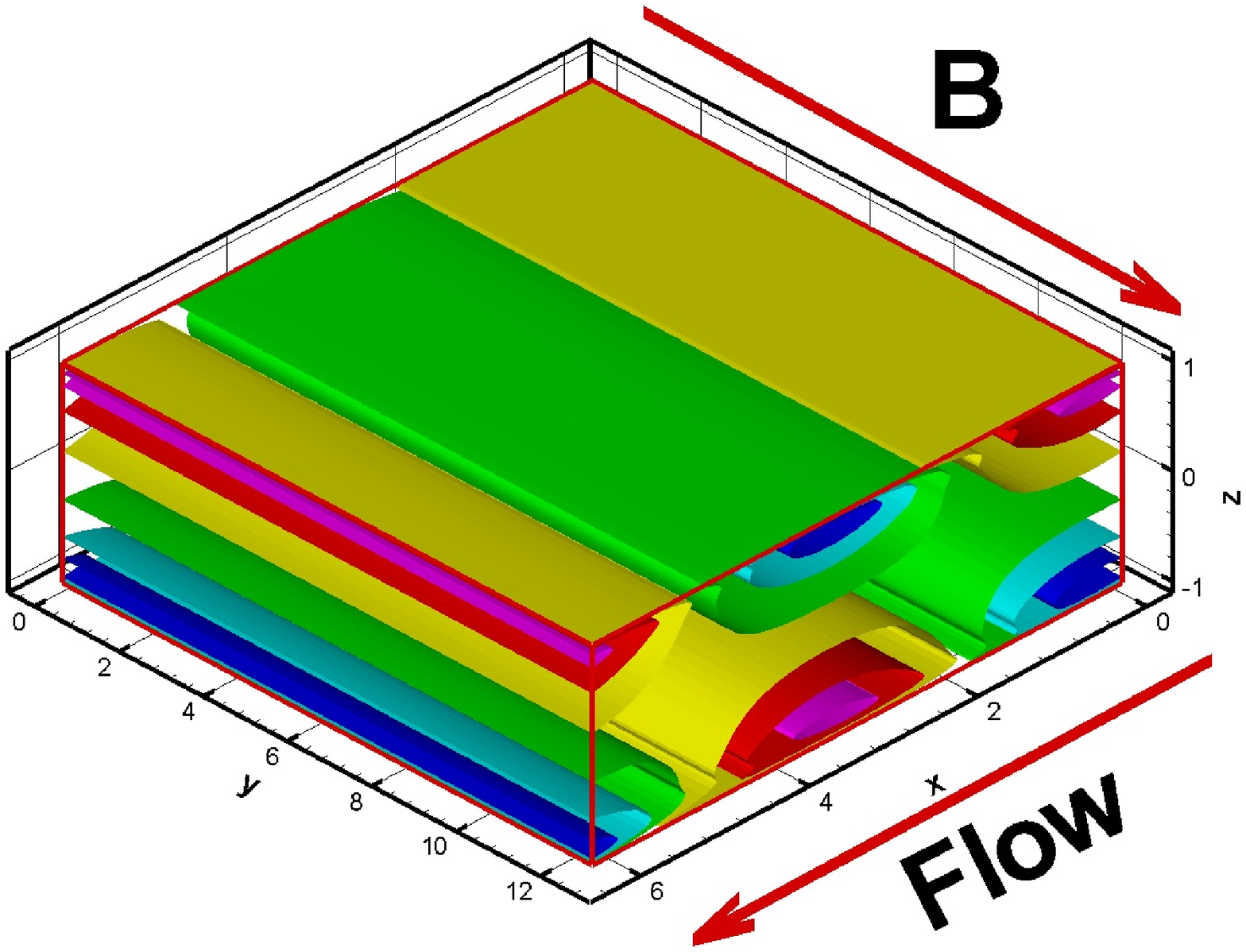}
\includegraphics[width=0.23\textwidth]{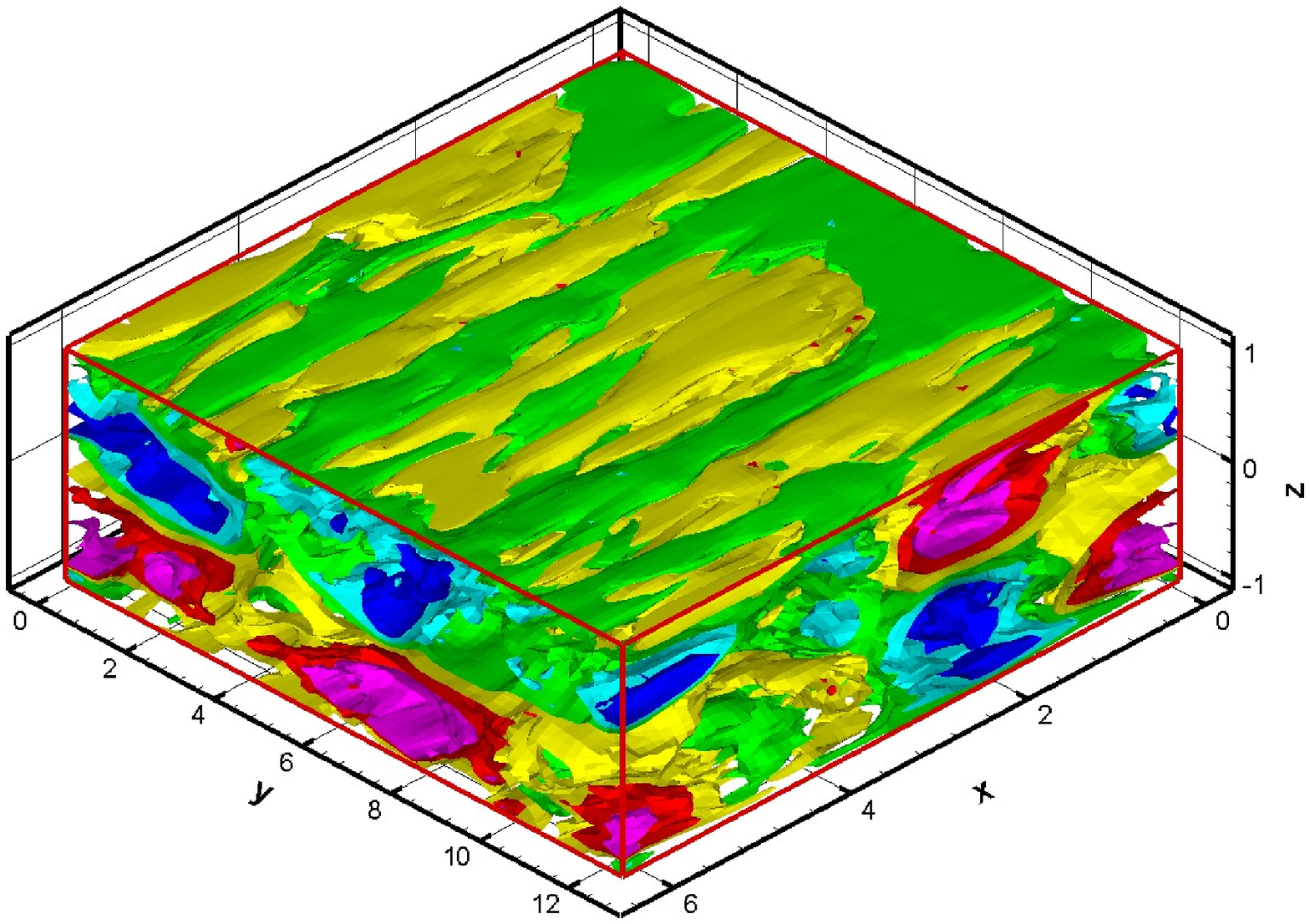}
}
\parbox{0.22\linewidth}{(c)}\parbox{0.76\linewidth}{(d)}
\centerline{
\includegraphics[width=0.23\textwidth]{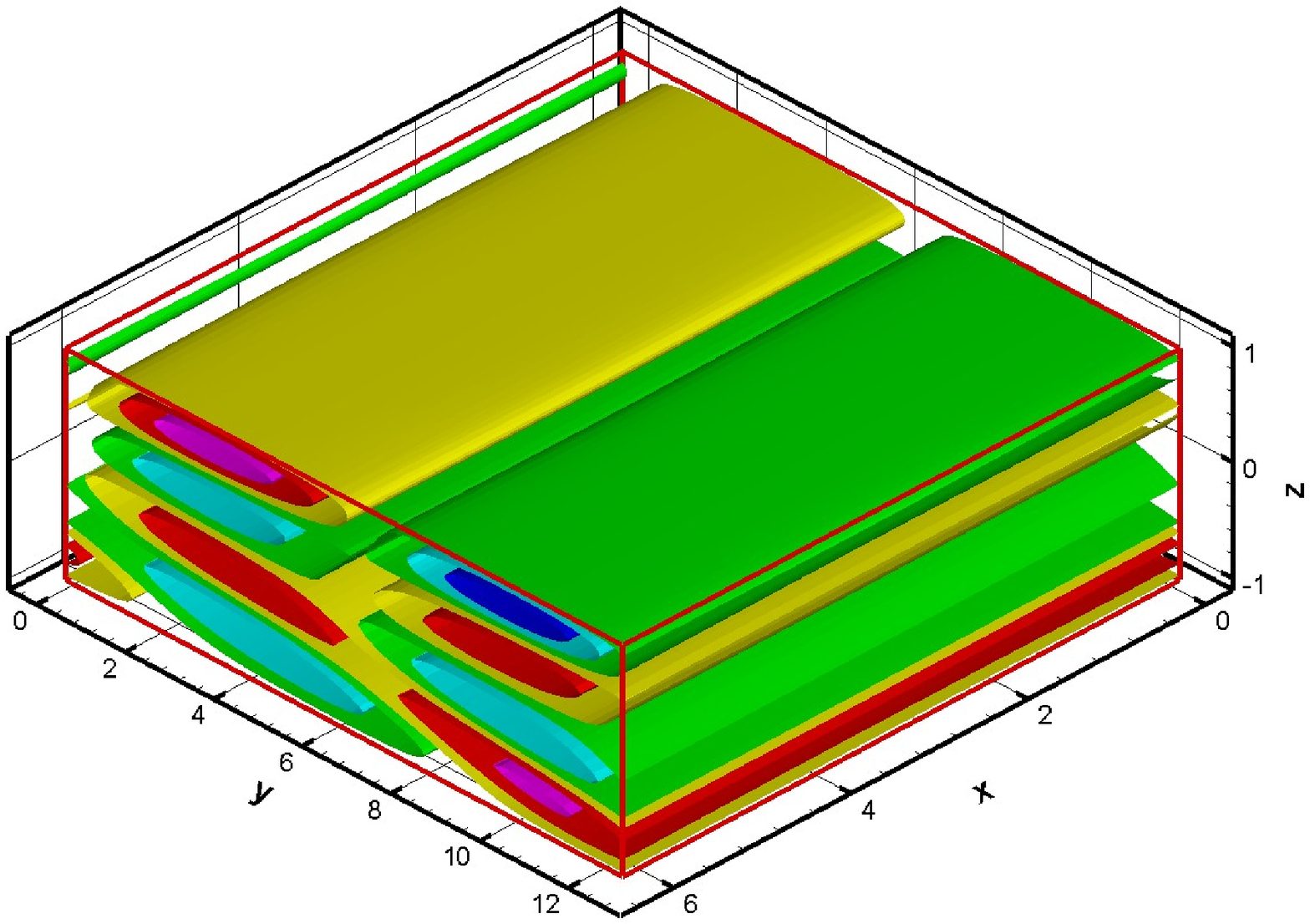}
\includegraphics[width=0.23\textwidth]{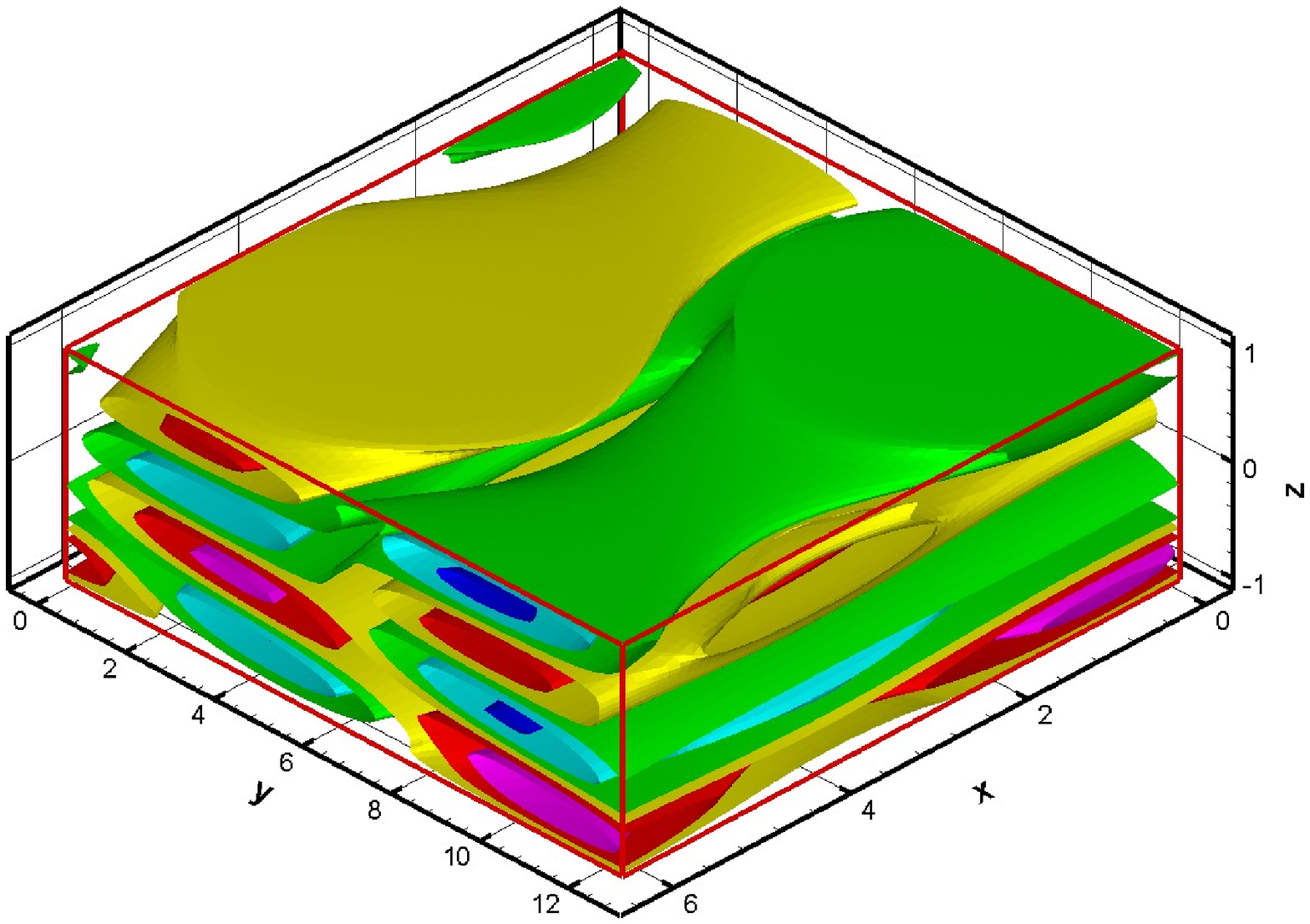}
} } \caption{\label{fig1:re08_080_zones}{Flow evolution during one
cycle of the intermittent process.
Four stages indicated in Fig.
\ref{fig1:re08_080_en} are shown using isosurfaces of streamwise
velocity perturbations normalized by corresponding rms-values. (a)
growing $2D$-spanwise TS mode, (b) $3D$ turbulent state 
at the maximum of perturbation energy,
(c) decaying flow dominated by streamwise streaks, (d)
disappearance of the streamwise streaks and return of
the growing TS waves. }}
\end{figure}

\begin{figure}[ht]
\centerline{
\includegraphics[width=0.45\textwidth]{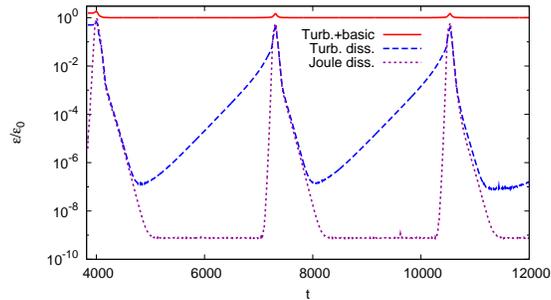}
} \caption{\label{fig1:re08_080_diss}{Evolution of viscous and
magnetic dissipation rates. 
 Total viscous dissipation (of
perturbations and basic flow), viscous dissipation of
perturbations only, and  Joule dissipation of perturbations are shown
(normalized to the dissipation $\epsilon_0$ of the basic flow.}}
\end{figure}

The flow transformation during one LSI cycle is presented in Figs.
\ref{fig1:re08_080_en}--\ref{fig1:re08_080_diss}.  Four stages can
be identified.  During the growth stage marked by \emph{(a)}, the
perturbation energy is almost exclusively in the spanwise-uniform
modes with wavenumber $k_y=0$.  It can be seen in Fig. \ref{fig1:re08_080_en}
that the energy of such modes shown by the short-dashed curve
constitutes nearly the entire energy of perturbations.  This conclusion
is confirmed by 2D energy power spectra $E(k_x,k_y)$
(not shown) and by the fact that the Joule dissipation rate shown
by long-dashed line in Fig. \ref{fig1:re08_080_diss} remains at
the level corresponding to dissipation of added noise.
Flow fields, visualized by the streamwise velocity component in
Fig. \ref{fig1:re08_080_zones}a, indicate that the growth phase
is dominated by the classical TS mode, i.e. the exponentially
growing solution of the linear stability problem of the 
basic flow. The growth rate $0.010752$ measured from the DNS results (branch
(a) in Fig.\ref{fig1:re08_080_en}) agrees with $0.010976$ obtained using
the linear stability code \cite{Krasnov:2007}.

After reaching finite amplitudes, the TS mode undergoes secondary
instability to 3D perturbations and  disintegrates to form a
turbulent state illustrated in Fig. \ref{fig1:re08_080_zones}b. The
identification of the state as turbulent is  supported by quick
population of the available $k_x$ and $k_y$ wavenumbers in the energy power
spectrum. The Joule dissipation rate increases sharply starting at
the moment of the first 3D instability of the TS mode and eventually
becomes comparable with the rate of viscous dissipation.  This
leads to strong suppression of the energy of perturbations and
initiates the stage of decay.

The decay stage marked by \emph{(c)} in Figs.
\ref{fig1:re08_080_en}--\ref{fig1:re08_080_zones} is characterized by
counterintuitive and somewhat surprising behavior.  Considering the
nature of Joule dissipation, one could expect that the spanwise
TS modes unaffected by the magnetic field would survive the
suppression.  This does not happen. 
A pattern of streamwise streaks illustrated in 
Fig. \ref{fig1:re08_080_zones}c 
develops as a dominant feature of the velocity perturbation 
field during the decay stage. The  nearly streamwise-independent
character of the flow is illustrated in Fig. \ref{fig1:re08_080_en},
where the short-dashed curve corresponding to the energy of purely
streamwise (with $k_x=0$) perturbations practically coincides with
the curve of the total perturbation energy.  The conclusion is also
supported by the 2D energy spectra.  The flow organization as a
system of streaks, i.e. zones of enhanced or reduced streamwise
velocity is visible in Fig. \ref{fig1:re08_080_zones}c and confirmed
by the fact that during this stage the energy of the streamwise
velocity component $\langle u'^2\rangle$ is at least two orders of
magnitude larger than the energy of the spanwise and wall-normal
components.

We only have a simplistic  explanation of the
dominance of streamwise streaks during the decay stage.
It is based  on the presence of coherent and relatively strong
streamwise streaks as a universal feature of turbulent channel
flow and, in general, of turbulence with mean shear.  It was shown
in our recent simulations \cite{Krasnov:2008} that this feature
persists in the presence of a moderate spanwise magnetic field (e.g.,
at $Re=10000$ and $Ha=30$). Moreover, the magnetic field renders the
streaks more coherent and somewhat larger in size  in all three
directions by suppressing small-scale 3D perturbations. Visual
indication of existence of streaks in the turbulent phase of the
intermittent flow can be seen in Fig. \ref{fig1:re08_080_zones}b.
We can assume that  spanwise TS modes are completely destroyed
in the turbulent flow, their energy being drained by instabilities into 
3D perturbations,  while the streaks form. As the 3D
fluctuations are suppressed by the magnetic field, the streaks of
largest spanwise wavelength (see Fig. \ref{fig1:re08_080_zones}c)
survive as least susceptible to Joule dissipation.

\begin{figure}[ht]
\centerline{
\includegraphics[width=0.45\textwidth]{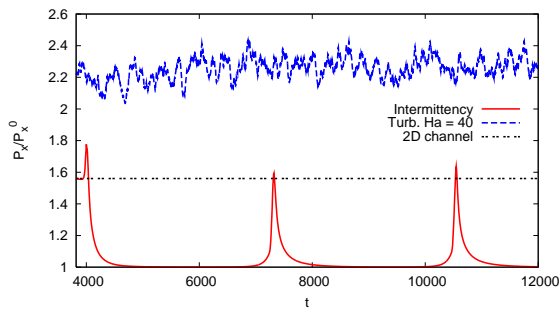}
} \caption{\label{fig1:re08_080_pgrad}{Mean pressure gradient
(normalized to the basic flow) at $Re=8000$
for the intermittent 
state at $Ha=80$,
for purely $2D$ channel flow \cite{Jimenez:1990}, and for fully developed
turbulence at  $Ha=40$.}}
\end{figure}

While the  streamwise streaks dominate the decaying perturbation
field, new spanwise modes form from the background noise. They
prepare the last stage of the intermittency cycle marked by
\emph{(d)} in Figs. \ref{fig1:re08_080_en} and
\ref{fig1:re08_080_zones}. It separates the decay and
growth phases and is characterized by comparable  energies 
of the growing TS mode and the  decaying streaks.
The total perturbation energy and the rate
of viscous dissipation assume the lowest values during this stage.
A change in  noise level affects the  
initial amplitude of the TS modes, whereby the decay  phase
of the streaks and the growth phase of the TS mode will  
be correspondingly shortened or lenghtened. 
The typical duration of the
LSI cycle was longer when  round-off errors were the only noise
source in our simulations. 

For a long fraction of the LSI cycle 
the flow remains close to the Poiseuille flow, which provides
the lowest friction drag 
for hydrodynamic
channel flow.  The drag experienced by 3D turbulent flow 
and purely 2D flow realized at lower and at higher Ha, respectively, 
are both on average 
substantially higher than for the  LSI. 
This non-monotonous drag reduction 
can be seen in Fig. \ref{fig1:re08_080_pgrad}, which shows
the mean pressure gradient $P_X$ needed to drive the flow. 

The spanwise domain size $L_y$ 
could have a potentially significant effect on the LSI.
Flow structures with longer spanwise wavelength
experience weaker Joule disspiation, and should
therefore persist up to larger values of $Ha$. The threshold for
LSI could therefore be shifted to higher $Ha$ when $L_y$
is increased. This question and the asymptotic behavior in the 
limit of very large $L_y$ could be partly addressed by a theoretical study of 
secondary instabilities of growing TS modes and of decaying streamwise streaks. 
Our present DNS approach alone  cannot
provide a satisfactory answer.
An experimental verification 
of the LSI could be attempted with
the low-melting eutectic alloy In-Ga-Sn, where $Re=8000$ and $Ha=80$ 
would correspond to $U\sim 1\textrm{m/s}$ and $B\sim 0.3 \textrm{T}$
for a channel with $L=1\textrm{cm}$. However,  the 
rigid lateral walls in a real channel flow 
present an important and yet undetermined factor. 

We are grateful to A. Tsinober for drawing our attention to
the problem of channel flow under spanwise magnetic field,
and to M. Rossi and J. Schumacher for  useful comments.
OZ is thankful to the Deutsche Forschungsgemeinschaft (DFG) for
support of his sabbatical stay at TU Ilmenau in the framework of
the Gerhard-Mercator visiting professorship program. TB, DK and AT
acknowledge financial support from the DFG in the framework of
the Emmy--Noether Program (grant Bo 1668/2-3).
OZ's work is supported by the grant DE FG02 03 ER46062 from
the U.S. Department of Energy. Computer resources were provided
by the computing centers of TU Ilmenau and of the Forschungszentrum
J\"ulich (NIC).


\end{document}